\documentclass[twoside]{article}
\usepackage{fleqn,espcrc2}
\usepackage{amsmath}

\usepackage[dvips]{graphicx}

\newcommand{\etal}{{\it et al.}}

\newcommand{\be}{\begin{eqnarray}}
\newcommand{\ee}{\end{eqnarray}}
\newcommand{\bdm}{\begin{displaymath}}
\newcommand{\edm}{\end{displaymath}}
\newcommand{\<}{\langle}
\renewcommand{\>}{\rangle}

\def\spose#1{\hbox to 0pt{#1\hss}}
\def\ltapprox{\mathrel{\spose{\lower 3pt\hbox{$\mathchar"218$}}
 \raise 2.0pt\hbox{$\mathchar"13C$}}}
\def\gtapprox{\mathrel{\spose{\lower 3pt\hbox{$\mathchar"218$}}
 \raise 2.0pt\hbox{$\mathchar"13E$}}}
\def\inapprox{\mathrel{\spose{\lower 3pt\hbox{$\mathchar"218$}}
 \raise 2.0pt\hbox{$\mathchar"232$}}}

\title{Thermodynamics with 3 and 2+1 Flavors of Improved Staggered Quarks
\thanks{Preprint FSU-CSIT-01-52. Talk given by U.M.~Heller at
{\it Lattice 2001}, August 19--24, Berlin, Germany.}}
\author{ C.~Bernard
\address{Department of Physics, Washington University, St.~Louis, MO 63130,
USA},
T.~Burch
\address{Department of Physics, University of Arizona, Tucson, AZ 85721, USA}, 
S.~Datta
\address{Department of Physics, Indiana University, Bloomington, IN 47405,
USA},
T.A.~DeGrand
\address{Physics Department, University of Colorado, Boulder, CO 80309, USA},
C.E.~DeTar
\address{Physics Department, University of Utah, Salt Lake City, UT
  84112, USA},
Steven~Gottlieb$\,\null^{\rm c}$,
U.M.~Heller
\address{CSIT, Florida State University, Tallahassee, FL 32306-4120, USA},
K.~Orginos
\address{RIKEN-BNL Research Center,
Brookhaven National Laboratory, Upton, NY 11973-5000},
R.L.~Sugar
\address{Department of Physics, University of California, Santa Barbara,
CA 93106, USA},
and D.~Toussaint$\,\null^{\rm b}$
} 

\begin{document}

\begin{abstract}
We present preliminary results \cite{SQCD}
from exploring the phase diagram of finite
temperature QCD with three degenerate flavors and with two light flavors
and the mass of the third held approximately at the strange quark mass.
We use an order $\alpha_s^2 a^2, a^4$ Symanzik improved gauge action and
an order $\alpha_s a^2, a^4$ improved staggered quark action. The improved
staggered action leads to a dispersion relation with diminished lattice
artifacts, and hence better thermodynamic properties. It decreases the
flavor symmetry breaking of staggered quarks substantially, and we estimate
that at the transition temperature for an $N_t=8$ to $N_t=10$ lattice
{\em all} pions will be lighter than the lightest kaon. Preliminary results
on lattices with $N_t=4$, 6 and 8 are presented.
\end{abstract}

\maketitle 

With the Relativistic Heavy Ion Collider (RHIC) now producing data,
it has become even more important to understand the phase diagram
of QCD at finite temperature, and to determine properties of the
high temperature quark-gluon-plasma phase with confidence,
{\it i.e.}~with controlled lattice spacing errors.

It is fairly well established that QCD with two flavors of massless
quarks has a second order finite temperature, chiral symmetry restoring
phase transition. This transition is washed out as soon as the quarks
become massive. QCD with three flavors of massless quarks has a first
order finite temperature, chiral symmetry restoring phase transition,
which is stable for small quark masses. Not well known is
how large the quark masses can be until the phase transition turns second
order and then into a crossover, both for degenerate quarks and especially
for the physically relevant case of two light and one heavier strange quark.

In previous studies, the second question is particularly badly answered
due to the flavor symmetry breaking in Kogut-Susskind quarks, usually used
for this purpose: how can one study the influence of the strange quark
when most of the (non-Goldstone) pions are heavier than the (Goldstone)
kaon?

Adding a few terms to the conventional Kogut-Susskind action, namely
three-link, five-link and seven-link staples and a third-neighbor
coupling, removes all tree-level ${\cal O}(a^2)$ errors
\cite{NAIK,MILC_FATTEST,LEPAGE98}.
This ``Asqtad'' action shows improved flavor
and rotational symmetry \cite{MILC_FATTEST,MILC_Spec}, and, at least
in quenched QCD, good scaling properties \cite{IMP_SCALING}.

\begin{figure}  
\centerline{\includegraphics[width=2.5in]{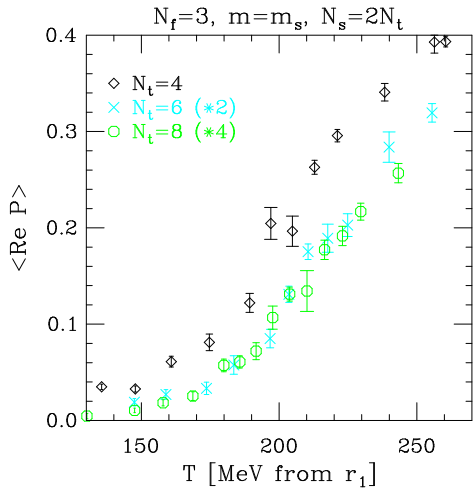}}
\vspace{-7mm}
\caption{Real part of the Polyakov line for three flavors with
$m_q \approx m_s$.
\label{fig:polr_m05} }
\vspace{-4mm}
\end{figure}

Based on our zero temperature simulations \cite{MILC_Spec} and the estimate
of $T_c \sim 150 - 170$ MeV \cite{FK_Schl01} we deduce that for $N_t=8 - 10$
the kaon will be heavier than the heaviest non-Goldstone pion at the finite
temperature transition.

We have zero temperature results, in particular, the value of the (bare) strange
quark mass, at fixed lattice spacing $a \sim 0.13$ fm and $a \sim 0.2$ fm.
Since we want to keep the physical quark masses approximately constant
when we vary the temperature (which, at fixed $N_t=1/(aT)$, means varying the
gauge coupling $\beta$), we interpolated (extrapolated) between the values
at the two lattice spacings.

\begin{figure}
\centerline{\includegraphics[width=2.5in]{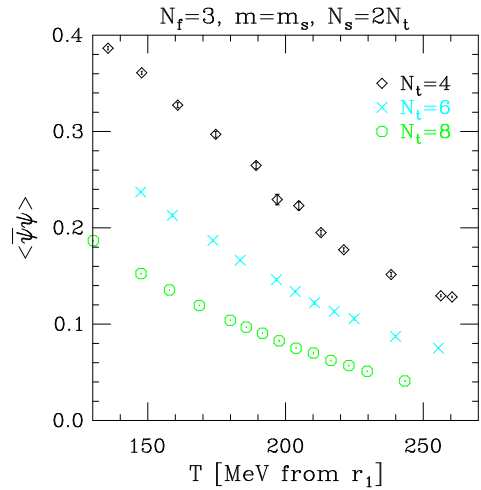}}
\vspace{-7mm}
\caption{$\< \bar \psi \psi \>$ for three flavors with $m_q \approx m_s$.
\label{fig:pbp_m05} }
\vspace{-4mm}
\end{figure}

We show in Fig.~\ref{fig:polr_m05} the real part of the Polyakov line
and in Fig.~\ref{fig:pbp_m05} the condensate as function of the
temperature for our simulations with three degenerate quarks of mass
$m_q \approx m_s$. To set the scale we used the distance $r_1$, defined
in terms of the static ${\rm Q\bar Q}$ potential by
$r_1^2 F_{\rm Q\bar Q static}(r_1) = 1$, {\it i.e.} $r_1 \sim 0.35$ fm
\cite{MILC_POT} and interpolated $a/r_1$, with the form advocated by
Allton~\cite{Allton}, between the two lattice spacings $a \sim 0.13$ fm
and $a \sim 0.2$ fm.

\begin{figure}
\centerline{\includegraphics[width=2.5in]{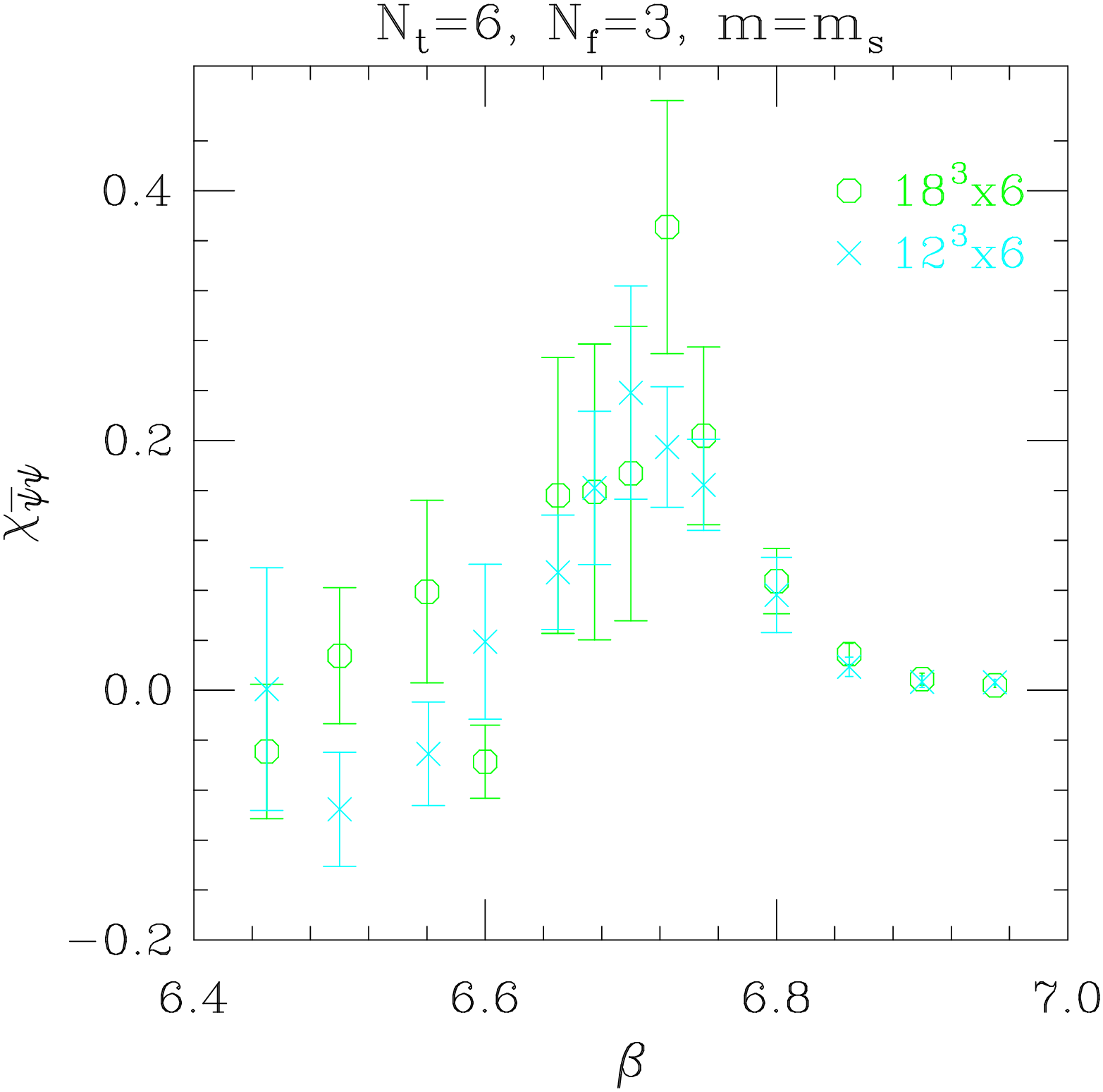}}
\vspace{-7mm}
\caption{The $\bar\psi \psi$ susceptibility for three flavors with
$m_q \approx m_s$ for lattices with $N_t=6$.
\label{fig:chi_t6_m05} }
\vspace*{3mm}
\centerline{\includegraphics[width=2.5in]{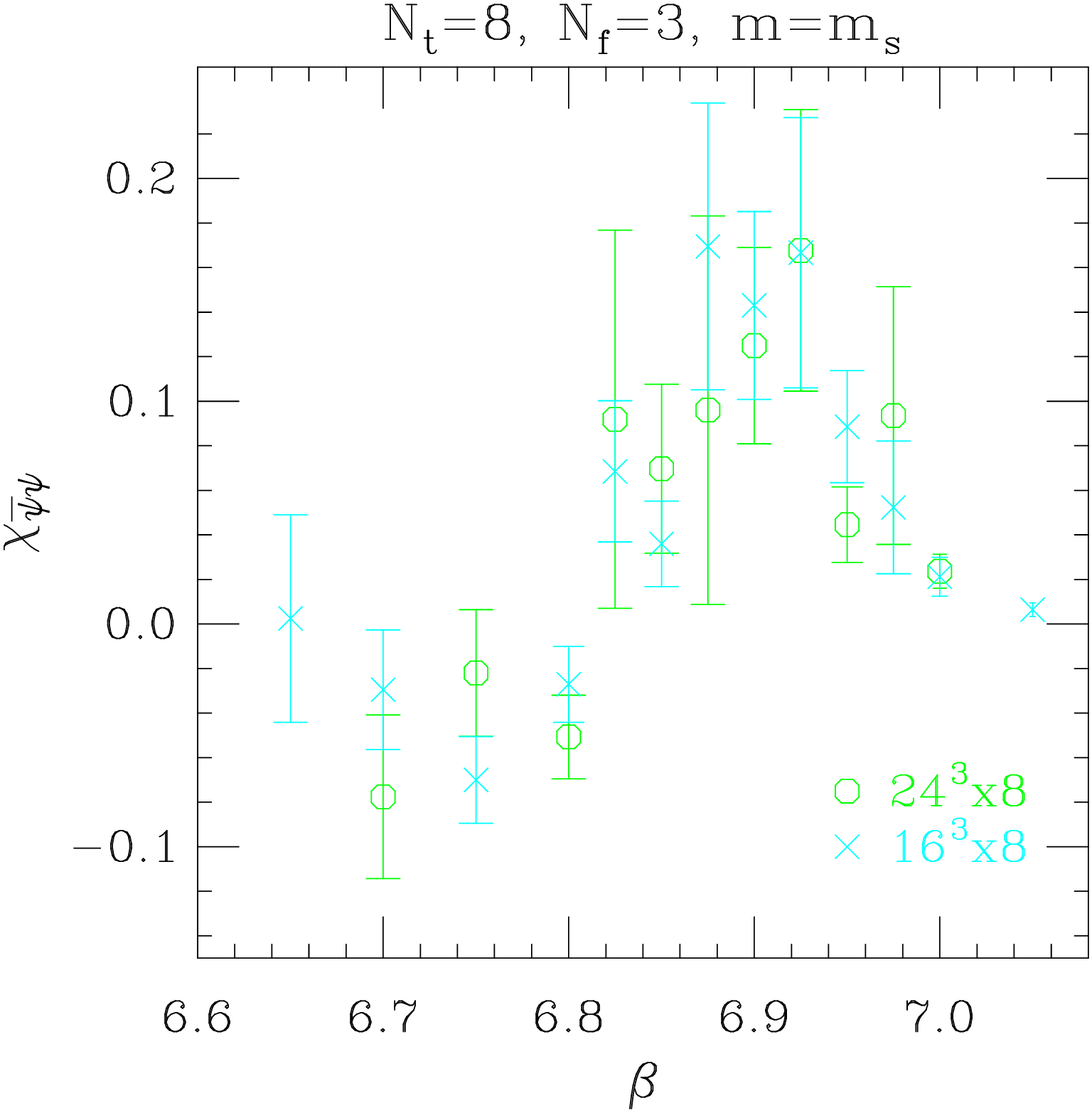}}
\vspace*{-7mm}
\caption{The $\bar\psi \psi$ susceptibility for three flavors with
$m_q \approx m_s$ for lattices with $N_t=8$.
\label{fig:chi_t8_m05} }
\vspace{-4mm}
\end{figure}

The Polyakov line shows a crossover behavior, but the condensate appears
to decrease, with increasing temperature, rather smoothly. However, the
chiral susceptibility, shown in Fig.~\ref{fig:chi_t6_m05} for $N_t=6$ and
in Fig.~\ref{fig:chi_t8_m05} for $N_t=8$, exhibits a peak. The peak height
appears to be independent of the spatial volume, indicative of a
smooth crossover.

\begin{figure}
\centerline{\includegraphics[width=2.5in]{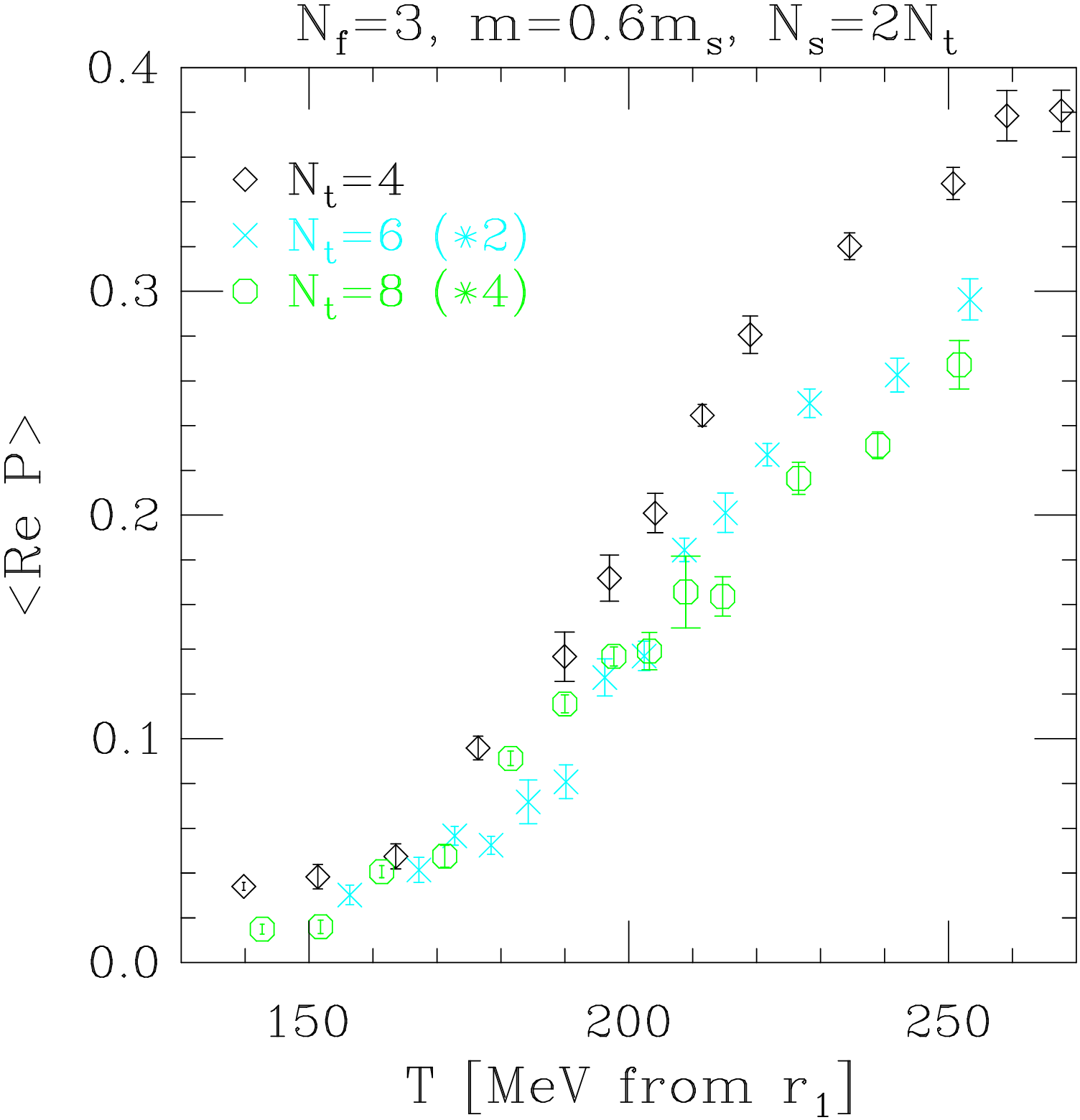}}
\vspace{-7mm}
\caption{Real part of the Polyakov line for three flavors with
$m_q \approx 0.6 m_s$.
\label{fig:polr_m03} }
\vspace*{3mm}
\centerline{\includegraphics[width=2.5in]{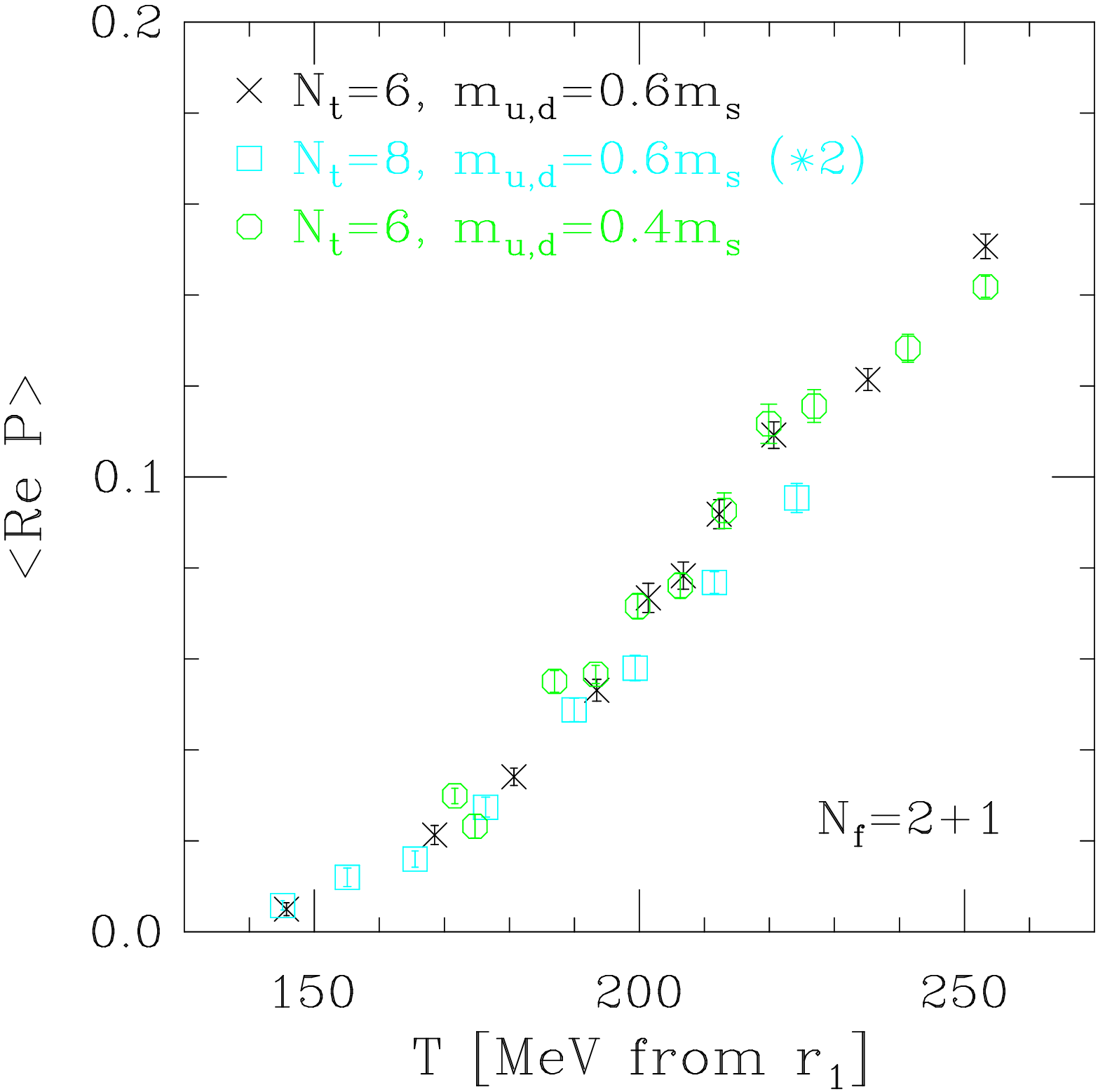}}
\vspace{-7mm}
\caption{Real part of the Polyakov line for 2+1 flavors.
\label{fig:polr_f21} }
\vspace{-4mm}
\end{figure}

Our other results look similar \cite{SQCD}. Fig.~\ref{fig:polr_m03} shows the
real part of the Polyakov line for our three-flavor simulations with
$m_q \approx 0.6 m_s$, and Fig.~\ref{fig:polr_f21} for our 2+1 flavor
simulations with $m_{u,d} = 0.6 m_s$ and $0.4 m_s$.

For the masses considered so far, we have not seen any sign of a genuine
phase transition. This result is compatible with other recent simulations
\cite{KS_fT} which found phase transitions only at quark masses lighter
than those studied by us so far. The temperature
at the crossover, with the scale set by $r_1$, $T_c \sim 190$ -- 200 MeV,
is a little higher than expected from previous determinations \cite{FK_Schl01}.
This is presumably also due to our quark masses still being rather high.

This work is supported by the US National Science Foundation and
Department of Energy and used computer resources at Florida State
University (SP), NERSC, NPACI, FNAL, and the University of Utah (CHPC).


\begin{thebibliography}{99}

\bibitem{SQCD}
For a more extended version, see
C.~Bernard \etal \, (The MILC collaboration), hep-lat/0110030.

\bibitem{NAIK}
S. Naik, Nucl. Phys. {\bf B316} (1989) 238.

\bibitem{MILC_FATTEST}
K. Orginos, D. Toussaint and R.L. Sugar,
Phys.\ Rev.\ D {\bf 60} (1999) 054503;
Nucl.\ Phys.\ (Proc.\ Suppl.)  {\bf 83} (2000) 878.

\bibitem{LEPAGE98}
G.P. Lepage, Phys.\ Rev.\ D {\bf 59} (1999) 074501.

\bibitem{MILC_Spec}
C.~Bernard \etal \, (The MILC collaboration),
Phys.\ Rev.\ D {\bf 64}, (2001) 054506.

\bibitem{IMP_SCALING}
C.~Bernard \etal \, (The MILC collaboration),
Phys.\ Rev.\ D {\bf 61}, (2000) 111502.

\bibitem{FK_Schl01}
See, {\it e.g.}, S. Ejiri, Nucl.\ Phys.\ B (Proc. Suppl) {\bf 94}, (2001) 19;
F. Karsch, hep-lat/0106019.

\bibitem{MILC_POT}
C.~Bernard \etal \, (The MILC collaboration),
Phys.\ Rev.\ D {\bf 62}, (2000) 034503.

\bibitem{Allton}
C. Allton, hep-lat/9610016.

\bibitem{KS_fT}
S. Aoki \etal \, (JLQCD), Nucl.\ Phys.\ B (Proc. Suppl) {\bf 73}, (1999) 459;
F. Karsch, E. Laermann and Ch. Schmidt, hep-lat/0107020.

\end{thebibliography}
\end{document}